\documentclass[10pt]{article}

\usepackage[english]{babel}
\usepackage[utf8x]{inputenc}
\usepackage[T1]{fontenc}
\usepackage{graphicx}
\usepackage{amsmath}
\usepackage{hyperref}
\usepackage{amssymb}
\usepackage[margin=1in, paperwidth=8.3in, paperheight=11.7in]{geometry}  
\usepackage{listings}
\title{How to solve kSAT in polynomial time }
\author{Maknickas Algirdas Antano}

\begin{document}
\maketitle 
\abstract{With using of multi-nary logic analytic formulas proposition that \textit{kSAT is in P and could be solved in $O\left(n^{3.5}\right)$} was proved.}

\section*{Introduction}

The Boolean satisfiability (SAT) problem~\cite{bib:Cook} is defined
as follows: Given a Boolean formula, check whether an
assignment of Boolean values to the propositional variables
in the formula exists, such that the formula evaluates to
true. If such an assignment exists, the formula is said to be
satisfiable; otherwise, it is unsatisfiable. For a formula with
m variables, there are 2m possible truth assignments. The
conjunctive normal form (CNF) 
\begin{equation}
(X_1 \vee X_2) \wedge (X_3 \vee X_4) \wedge \dots \wedge (X_{n-1} \vee X_n)
\end{equation}
is most the frequently used
for representing Boolean formulas, where $\neg \forall X_i$ are independent. In CNF, the variables
of the formula appear in literals (e.g., x) or their negation
(e.g., $\neg$x (logical NOT $\neg$)). Literals are grouped into clauses, which represent a
disjunction (logical OR $\vee$) of the literals they contain. A single
literal can appear in any number of clauses. The conjunction
(logical AND $\wedge$) of all clauses represents a formula.  

Several algorithms are known for solving the 2-satisfiability problem; the most efficient of them take linear time~\cite{bib:Krom},~\cite{bib:Aspvall},~\cite{bib:Even}. Instances of the 2-satisfiability or 2-SAT problem are typically expressed as 2-CNF or Krom formulas~\cite{bib:Krom}

SAT was the first known NP-complete problem, as proved by Cook and Levin in 1971 ~\cite{bib:Cook}~\cite{bib:Cook-Levin}. Until that time, the concept of an NP-complete problem did not even exist. The problem remains NP-complete even if all expressions are written in conjunctive normal form with 3 variables per clause (3-CNF), yielding the 3SAT problem. This means the expression has the form:
\begin{equation}
(X_1 \vee X_2 \vee X_3) \wedge (X_4 \vee X_5 \vee X_6) \wedge \dots \wedge (X_{n-2} \vee X_{n-1} \vee X_n)
\end{equation}
NP-complete and it is used as a starting point for proving that other problems are also NP-hard. This is done by polynomial-time reduction from 3-SAT to the other problem.  

Fagin formulated in their article ~\cite{bib:Fagin}, that the following two statements are equivalent: $NP = P$ and  \textit{There exists a constant $k$ such that, for every countable function $T$
with $T\left(l\right) \geq l + 1$ for each l and for every language $A$ which is recognized
by a non-deterministic one-tape Turing machine in time $T$, the language $A$ is
recognized by a deterministic one-tape Turing machine in time $T^k$}  Author of this article proposed proof of this theorem in ~\cite{bib:Maknickas}. After this publication Weiss proposed A Polynomial Algorithm for 3-sat~\cite{bib:Weiss}. Sergey Kardash~\cite{bib:Kardash} described  polynomial algorithm for solving k-satisfiability ($k \ge 2$) problem and stated that each problem from NP can be solved polynomially or P=NP.  Matt Groff~\cite{bib:Groff} established P=NP through an $O(n^7)$ time algorithm for the satisfiability problem. All this proofs are not reviewed and accepted of public mathematical society until now.

The goal of this paper is proof of proposition that kSAT is in P using multi logic formulas of discrete second order logic proposed first in~\cite{bib:Maknickas}. 
\section{Multi-nary logic formulas in modulo form}
Formulas given in ~\cite{bib:Maknickas} could be expressed in modulo notations $a_i = b_i \pmod n $ for integers.
Let describe integer discrete logic units as $i$, where $i \in \mathbb{Z}^+ $. 
Let describe discrete function $g_k^n\left(a\right), \forall a \in \mathbb{R}, \forall k \in \left\lbrace 0, 1, 2, ... , n-1\right\rbrace$ as 
\begin{equation}
g_k^n\left(a\right) = {\lfloor a \rfloor + k} \pmod n
\end{equation}

LEMMA 1. \textit{If $n = 2$, function $g_k^n\left(a\right)$ is one variable binary logic generation function for binary set $ \left\lbrace 0, 1\right\rbrace $, where $0$ is \text{true} and $1$ is \text{false}}.\\
\textit{Proof}. The are $2^2$ different one variable logic functions:

\begin{align}
\varrho^{i_0}_{i_1}\left(a\right) &=
 \begin{array}{c|c}
  \lfloor a \rfloor & rez \\  
  \hline  
  \mbox{$0$} & \mbox{$g^2_{i_0}\left(a\right)$} \\
  \mbox{$1$} & \mbox{$g^2_{i_1}\left(a\right)$}
 \end{array}, \text{\ \ } \forall i_0, i_1 \in \left\lbrace 0, 1 \right\rbrace 
 \nonumber \\
\varrho^{0}_{0}\left(a\right) &=
 \begin{array}{c|c}
  \lfloor a \rfloor & rez \\  
  \hline  
  \mbox{$0$} & \mbox{$g^2_0\left(a\right)$} \\
  \mbox{$1$} & \mbox{$g^2_0\left(a\right)$}
 \end{array}
 =
  \begin{array}{c|c}
  \lfloor a \rfloor & rez \\  
  \hline  
  \mbox{$0$} & \mbox{$0$} \\
  \mbox{$1$} & \mbox{$1$}
 \end{array} \nonumber \\
\varrho^{0}_{1}\left(a\right) &=
\begin{array}{c|c}
  \lfloor a \rfloor & rez \\  
  \hline  
  \mbox{$0$} & \mbox{$g^2_0\left(a\right)$} \\
  \mbox{$1$} & \mbox{$g^2_1\left(a\right)$}
 \end{array}
 =
  \begin{array}{c|c}
  \lfloor a \rfloor & rez \\  
  \hline  
  \mbox{$0$} & \mbox{$0$} \\
  \mbox{$1$} & \mbox{$0$}
 \end{array} \label{eq:5} \\
\varrho^{1}_{0}\left(a\right) &= 
 \begin{array}{c|c}
  \lfloor a \rfloor & rez \\  
  \hline  
  \mbox{$0$} & \mbox{$g^2_1\left(a\right)$} \\
  \mbox{$1$} & \mbox{$g^2_0\left(a\right)$}
 \end{array}
 =
  \begin{array}{c|c}
  \lfloor a \rfloor & rez \\  
  \hline  
  \mbox{$0$} & \mbox{$1$} \\
  \mbox{$1$} & \mbox{$1$}
 \end{array} \nonumber \\
\varrho^{1}_{1}\left(a\right) &= 
 \begin{array}{c|c}
  \lfloor a \rfloor & rez \\  
  \hline  
  \mbox{$0$} & \mbox{$g^2_1\left(a\right)$} \\
  \mbox{$1$} & \mbox{$g^2_1\left(a\right)$}
 \end{array}
 =
  \begin{array}{c|c}
  \lfloor a \rfloor & rez \\  
  \hline  
  \mbox{$0$} & \mbox{$1$} \\
  \mbox{$1$} & \mbox{$0$}
 \end{array} \nonumber
 \end{align}
 
Direct calculations show, that $\varrho^{0}_{0}$ is self projection, $\varrho^{0}_{1}$ is antilogy, $\varrho^{1}_{0}$ is tautology, $\varrho^{1}_{1}$ is complementation.  $\bigcirc$ 
 
LEMMA 2. \textit{If $n = 2$, function $g_k^n\left(a*b\right)$ is two variables binary logic generation function for binary set $\left\lbrace 0, 1\right\rbrace $, where $0$ names \textit{true} and $1$ names \textit{false}}. 

\textit{Proof}. The are $2^{2^2}$ different two variables logic functions:

 \begin{align}
\mu^{i_0,i_1}_{i_2,i_3}\left( a, b \right)
&=  
  \begin{array}{c|c c}
  \lfloor a \rfloor \backslash \lfloor b \rfloor &  \mbox{$0$} & \mbox{$1$}\\  
  \hline  
  \mbox{$0$} & \mbox{$g_{i_0}\left( a*b \right)$} & \mbox{$g_{i_1}\left( a*b \right)$} \\
  \mbox{$1$} & \mbox{$g_{i_2}\left( a*b \right)$} & \mbox{$g_{i_3}\left( a*b \right)$} 
 \end{array}, \text{\ \ } \forall i_0, i_1, i_2, i_3 \in \left\lbrace 0, 1 \right\rbrace  \nonumber 
\end{align}
\begin{align}
 \mu^{0,0}_{0,0}\left( a, b \right)
&=  
  \begin{array}{c|c c}
  \lfloor a \rfloor \backslash \lfloor b \rfloor &  \mbox{$0$} & \mbox{$1$}\\  
  \hline  
  \mbox{$0$} & \mbox{$0$} & \mbox{$0$} \\
  \mbox{$1$} & \mbox{$0$} & \mbox{$1$} 
 \end{array}, 
  \mu^{0,0}_{0,1}\left( a, b \right)
=  
  \begin{array}{c|c c}
  \lfloor a \rfloor \backslash \lfloor b \rfloor &  \mbox{$0$} & \mbox{$1$}\\  
  \hline  
  \mbox{$0$} & \mbox{$0$} & \mbox{$0$} \\
  \mbox{$1$} & \mbox{$0$} & \mbox{$0$} 
 \end{array} \nonumber 
\end{align}
\begin{align}
  \mu^{0,0}_{1,0}\left( a, b \right)
&=  
  \begin{array}{c|c c}
  \lfloor a \rfloor \backslash \lfloor b \rfloor &  \mbox{$0$} & \mbox{$1$}\\  
  \hline  
  \mbox{$0$} & \mbox{$0$} & \mbox{$0$} \\
  \mbox{$1$} & \mbox{$1$} & \mbox{$1$} 
 \end{array}, 
  \mu^{0,0}_{1,1}\left( a, b \right)
=  
  \begin{array}{c|c c}
  \lfloor a \rfloor \backslash \lfloor b \rfloor &  \mbox{$0$} & \mbox{$1$}\\  
  \hline  
  \mbox{$0$} & \mbox{$0$} & \mbox{$0$} \\
  \mbox{$1$} & \mbox{$1$} & \mbox{$0$} 
 \end{array} \nonumber 
\end{align}
\begin{align}
  \mu^{0,1}_{0,0}\left( a, b \right)
&=  
  \begin{array}{c|c c}
  \lfloor a \rfloor \backslash \lfloor b \rfloor &  \mbox{$0$} & \mbox{$1$}\\  
  \hline  
  \mbox{$0$} & \mbox{$0$} & \mbox{$1$} \\
  \mbox{$1$} & \mbox{$0$} & \mbox{$1$} 
 \end{array},  
  \mu^{0,1}_{0,1}\left( a, b \right)
=  
  \begin{array}{c|c c}
  \lfloor a \rfloor \backslash \lfloor b \rfloor &  \mbox{$0$} & \mbox{$1$}\\  
  \hline  
  \mbox{$0$} & \mbox{$0$} & \mbox{$1$} \\
  \mbox{$1$} & \mbox{$0$} & \mbox{$0$} 
 \end{array} \nonumber 
\end{align}
\begin{align}
  \mu^{0,1}_{1,0}\left( a, b \right)
&=  
  \begin{array}{c|c c}
  \lfloor a \rfloor \backslash \lfloor b \rfloor &  \mbox{$0$} & \mbox{$1$}\\  
  \hline  
  \mbox{$0$} & \mbox{$0$} & \mbox{$1$} \\
  \mbox{$1$} & \mbox{$1$} & \mbox{$1$} 
 \end{array},
  \mu^{0,1}_{1,1}\left( a, b \right)
=  
  \begin{array}{c|c c}
  \lfloor a \rfloor \backslash \lfloor b \rfloor &  \mbox{$0$} & \mbox{$1$}\\  
  \hline  
  \mbox{$0$} & \mbox{$0$} & \mbox{$1$} \\
  \mbox{$1$} & \mbox{$1$} & \mbox{$0$} 
 \end{array} \label{eq:6} 
\end{align}
\begin{align}
  \mu^{1,0}_{0,0}\left( a, b \right)
&=  
  \begin{array}{c|c c}
  \lfloor a \rfloor \backslash \lfloor b \rfloor &  \mbox{$0$} & \mbox{$1$}\\  
  \hline  
  \mbox{$0$} & \mbox{$1$} & \mbox{$0$} \\
  \mbox{$1$} & \mbox{$0$} & \mbox{$1$} 
 \end{array}, 
  \mu^{1,0}_{0,1}\left( a, b \right)
=  
  \begin{array}{c|c c}
  \lfloor a \rfloor \backslash \lfloor b \rfloor &  \mbox{$0$} & \mbox{$1$}\\  
  \hline  
  \mbox{$0$} & \mbox{$1$} & \mbox{$0$} \\
  \mbox{$1$} & \mbox{$0$} & \mbox{$0$} 
 \end{array} \nonumber 
\end{align}
\begin{align}
  \mu^{1,0}_{1,0}\left( a, b \right)
&=  
  \begin{array}{c|c c}
  \lfloor a \rfloor \backslash \lfloor b \rfloor &  \mbox{$0$} & \mbox{$1$}\\  
  \hline  
  \mbox{$0$} & \mbox{$1$} & \mbox{$0$} \\
  \mbox{$1$} & \mbox{$1$} & \mbox{$1$} 
 \end{array}, 
  \mu^{1,0}_{1,1}\left( a, b \right)
=  
  \begin{array}{c|c c}
  \lfloor a \rfloor \backslash \lfloor b \rfloor &  \mbox{$0$} & \mbox{$1$}\\  
  \hline  
  \mbox{$0$} & \mbox{$1$} & \mbox{$0$} \\
  \mbox{$1$} & \mbox{$1$} & \mbox{$0$} 
 \end{array} \nonumber 
\end{align}
\begin{align}
  \mu^{1,1}_{0,0}\left( a, b \right)
&=  
  \begin{array}{c|c c}
  \lfloor a \rfloor \backslash \lfloor b \rfloor &  \mbox{$0$} & \mbox{$1$}\\  
  \hline  
  \mbox{$0$} & \mbox{$1$} & \mbox{$1$} \\
  \mbox{$1$} & \mbox{$0$} & \mbox{$1$} 
 \end{array}, 
  \mu^{1,1}_{0,1}\left( a, b \right)
=  
  \begin{array}{c|c c}
  \lfloor a \rfloor \backslash \lfloor b \rfloor &  \mbox{$0$} & \mbox{$1$}\\  
  \hline  
  \mbox{$0$} & \mbox{$1$} & \mbox{$1$} \\
  \mbox{$1$} & \mbox{$0$} & \mbox{$0$} 
 \end{array} \nonumber 
\end{align}
\begin{align}
  \mu^{1,1}_{1,0}\left( a, b \right)
&=  
  \begin{array}{c|c c}
  \lfloor a \rfloor \backslash \lfloor b \rfloor &  \mbox{$0$} & \mbox{$1$}\\  
  \hline  
  \mbox{$0$} & \mbox{$1$} & \mbox{$1$} \\
  \mbox{$1$} & \mbox{$1$} & \mbox{$1$} 
 \end{array}, 
 \mu^{1,1}_{1,1}\left( a, b \right)
=  
  \begin{array}{c|c c}
  \lfloor a \rfloor \backslash \lfloor b \rfloor &  \mbox{$0$} & \mbox{$1$}\\  
  \hline  
  \mbox{$0$} & \mbox{$1$} & \mbox{$1$} \\
  \mbox{$1$} & \mbox{$1$} & \mbox{$0$} 
 \end{array} \nonumber 
 \end{align}
 
 Direct calculations show, that $\mu^{0,0}_{0,0}$ is nand, $\mu^{0,0}_{0,1}$ is antilogy, $\mu^{0,0}_{1,0}$ is left complementation, $\mu^{0,0}_{1,1}$ is if ... then, $\mu^{0,1}_{0,0}$ is right projection, $\mu^{0,1}_{0,1}$ is if, $\mu^{0,1}_{1,0}$ is neither ... nor, $\mu^{0,1}_{1,1}$ is if and only if (iff), $\mu^{1,0}_{0,0}$ is xor, $\mu^{1,0}_{0,1}$ is or, $\mu^{1,0}_{1,0}$ is not ... but, $\mu^{1,0}_{1,1}$ is right projection, $\mu^{1,1}_{0,0}$ is but not, $\mu^{1,1}_{0,1}$ is left projection, $\mu^{1,1}_{1,0}$ is tautology, $\mu^{1,1}_{1,1}$ is and \cite{bib:Knuth}. $\bigcirc$

LEMMA 3. \textit{If $n > 2$, function $g_k^n\left(a\right)$ is one variable multi-nary logic generation function for multi-nary set $ \left\lbrace 0, 1, 2, .. , {n-1} \right\rbrace $}

\textit{Proof}. The are $n^n$ one variable logic functions:

\begin{align}
\varrho \begin{array}{l}
         \mbox{$i_0$} \\
         \mbox{$i_1$} \\
         \mbox{$i_2$} \\
         \dots \\
         \mbox{$i_{n-1}$} \end{array} \left(a\right) &=
 \begin{array}{l|l}
  \lfloor a \rfloor & rez \\  
  \hline  
  \mbox{$0$} & \mbox{$g^n_{i_0}\left(a\right)$} \\
  \mbox{$1$} & \mbox{$g^n_{i_1}\left(a\right)$} \\
  \mbox{$2$} & \mbox{$g^n_{i_2}\left(a\right)$} \\
  \dots & \dots \\
  \mbox{${n-1}$} & \mbox{$g^n_{i_{n-1}}\left(a\right)$}
 \end{array}, \text{\ \ } \forall \begin{array}{l}
         \mbox{$i_0$} \\
         \mbox{$i_1$} \\
         \mbox{$i_2$} \\
         \dots \\
         \mbox{$i_{n-1}$} \end{array} \in \left\lbrace 0, 1, 2, ... , n-1 \right\rbrace  
\end{align}

All $\varrho$ function could be generated starting from index set $\left\{ i_0, i_1, i_2, ... , i_{n-1} \right\}$ $=$ $\left\lbrace 0, 0, 0, ... , 0 \right\rbrace$. For every two nearest $\varrho$ functions with index sets $\left\{ i_l, i_l, i_l,\right.$ $\left. ... , i_k, ... , i_l \right\}$ and $\left\{ i_l, i_l, i_l, ... , i_k+1, ... , i_l \right\}$ functions  $g^n_{i_l}\left(a\right) = g^n_{i_l} \left(a\right)$ and $g^n_{i_k}\left(a\right) \neq g^n_{i_k+1} \left(a\right)$. So all $n^n$ $\varrho$ functions with unique index set $\left\{ i_0, i_1, \right.$ $\left. i_2, ... , i_{n-1} \right\}$  are different. $\bigcirc$ 

LEMMA 4. \textit{If $n > 2$,  function $g_k^n\left(a*b\right)$ is two variables multi-nary logic generation function for multi-nary set $\left\lbrace 0, 1, 2, ... , {n-1}\right\rbrace $}. 

\textit{Proof}. The are $n^{n^2}$ two variables logic functions:

\begin{align}
\mu \begin{array}{lllll}
         \mbox{$i_{0,0}$} & \mbox{$i_{0,1}$} & \mbox{$i_{0,2}$} & \dots & \mbox{$i_{0, n-1}$} \\
         \mbox{$i_{1,0}$} & \mbox{$i_{1,1}$} & \mbox{$i_{1,2}$} & \dots & \mbox{$i_{1, n-1}$} \\
         \mbox{$i_{2,0}$} & \mbox{$i_{2,1}$} & \mbox{$i_{2,2}$} & \dots & \mbox{$i_{2, n-1}$} \\
         \dots & \dots & \dots & \dots & \dots \\
         \mbox{$i_{n-1, 0}$} & \mbox{$i_{n-1, 1}$} & \mbox{$i_{n-1, 2}$} & \dots & \mbox{$i_{n-1, n-1}$} \end{array}   &= \nonumber 
\end{align}
\begin{align} 
 \begin{array}{l|lllll}
  \lfloor a \rfloor \backslash \lfloor b \rfloor & \mbox{$0$} & \mbox{$1$} & \mbox{$2$} & \dots & \mbox{${n-1}$} \\  
  \hline  
  \mbox{$0$} & \mbox{$g^n_{i_{0,0}}\left({a*b}\right)$} & \mbox{$g^n_{i_{0,1}}\left({a*b}\right)$} & \mbox{$g^n_{i_{0,2}}\left({a*b}\right)$} & \dots & \mbox{$g^n_{i_{0,n-1}}\left({a*b}\right)$}\\
  \mbox{$1$} & \mbox{$g^n_{i_{1,0}}\left({a*b}\right)$} & \mbox{$g^n_{i_{1,1}}\left({a*b}\right)$} & \mbox{$g^n_{i_{1,2}}\left({a*b}\right)$} & \dots & \mbox{$g^n_{i_{1,n-1}}\left({a*b}\right)$} \\
  \mbox{$2$} & \mbox{$g^n_{i_{2,0}}\left({a*b}\right)$} & \mbox{$g^n_{i_{2,1}}\left({a*b}\right)$} & \mbox{$g^n_{i_{2,2}}\left({a*b}\right)$} & \dots & \mbox{$g^n_{i_{2,n-1}}\left({a*b}\right)$} \\
  \dots & \dots & \dots & \dots & \dots & \dots \\
  \mbox{${n-1}$} & \mbox{$g^n_{i_{n-1,0}}\left({a*b}\right)$} & \mbox{$g^n_{i_{n-1,1}}\left({a*b}\right)$} & \mbox{$g^n_{i_{n-1,2}}\left({a*b}\right)$} & \dots & \mbox{$g^n_{i_{n-1,n-1}}\left({a*b}\right)$}
 \end{array}, \nonumber
 \end{align}
 \begin{equation}
   \text{\ \ } \forall \begin{array}{lllll}
         \mbox{$i_{0,0}$} & \mbox{$i_{0,1}$} & \mbox{$i_{0,2}$} & \dots & \mbox{$i_{0, n-1}$} \\
         \mbox{$i_{1,0}$} & \mbox{$i_{1,1}$} & \mbox{$i_{1,2}$} & \dots & \mbox{$i_{1, n-1}$} \\
         \mbox{$i_{2,0}$} & \mbox{$i_{2,1}$} & \mbox{$i_{2,2}$} & \dots & \mbox{$i_{2, n-1}$} \\
         \dots & \dots & \dots & \dots & \dots \\
         \mbox{$i_{n-1, 0}$} & \mbox{$i_{n-1, 1}$} & \mbox{$i_{n-1, 2}$} & \dots & \mbox{$i_{n-1, n-1}$} \end{array} \in \left\lbrace 0, 1, 2, ... , n-1 \right\rbrace  
\end{equation}
All $\mu$ function could be generated starting from index set $\left\{ i_{0,0}, i_{0,1}, i_{0,2}, ... , i_{n-1,n-1} \right\} = \left\lbrace 0, 0, 0, ... , 0 \right\rbrace$. For every two nearest $\mu$ functions with index sets $\left\{ i_{l1,l2}, i_{l1,l2}, i_{l1,l2}, ... , i_{k1,k2}, ... , i_{l1,l2} \right\}$ \\and $\left\{ i_{l1,l2}, i_{l1,l2}, i_{l1,l2}, ... , i_{k1,k2}+1, ... , i_{l,l} \right\}$ functions $g^n_{i_{l1,l2}}\left({a*b}\right) = g^n_{i_{l1,l2}} \left({a*b}\right)$ and $g^n_{i_{k1,k2}}\left({a*b}\right) \neq g^n_{i_{k1,k2}+1} \left({a*b}\right)$. So all $n^{n^2}$ $\mu$ functions with unique index set $\left\{ i_{0,0}, i_{0,1}, i_{0,2}, ... , i_{n-1,n-1} \right\}$  are different. $\bigcirc$ 
\section{2SAT is in P}
\textit{THEOREM 1. If binary multi-variable logic function is expressed as}
\begin{equation}
\beta_2 \left(X_1, X_2, ... , X_n\right) = (X_1 \vee X_2) \wedge (X_3 \vee X_4) \wedge \dots \wedge (X_{n-1} \vee X_n),
\end{equation}   
\textit{it could be calculated in} $O\left(m\right)$ \textit{where $m$ is number of clauses and $m \ge n$}.

\textit{Proof}. Let start to investigate $\beta$. It could be expressed in notations of LEMMA 4 as
\begin{align}
\mu \left(a,b\right) &= 
  \begin{array}{c|c c c}
  \lfloor a \rfloor \backslash \lfloor b \rfloor &  \mbox{$0$} & \mbox{$1$} & \mbox{$2$}\\  
  \hline  
  \mbox{$0$} & \mbox{$0$} & \mbox{$0$} & \mbox{$0$}\\
  \mbox{$1$} & \mbox{$0$} & \mbox{$1$} & \mbox{$1$}\\
  \mbox{$2$} & \mbox{$0$} & \mbox{$1$} & \mbox{$1$}
 \end{array} \\
 \beta_2 \left(X_1, X_2, ..., X_{n-1} , X_n\right) &= \mu \left(X_1 + X_2, \mu \left( X_3 + X_4, ... , \mu \left( X_{n-3} + X_{n-2}, X_{n-1} + X_{n} \right) \right) \right) \label{eq:10} 
\end{align}
where $+$ is algebraic summation. So expressed $\beta$ function could be calculated  within $m/2$ summations and $m/2-1$ calls of $\mu$ function. Every result of $\mu$ function (getting item from 2 dimensional array ) could be calculated within $2$ summation operations (one for finding row and one for finding column). So total $\beta$ function calculation time could be expressed as
\begin{align}
T_{\sum} = t_{+} \frac{m}{2} + 2 t_{+} \left( \frac{m}{2} -1\right) + p t_{-}= O\left(m\right)
\end{align}  
where $t_+$ is algebraic summation time of two variables and is constant, $t_{-}$ is unary negotiation time and $p$ amount of negotiation functions $p \le n$. $\beta$ function could be tested in linear time. $\bigcirc$

\textit{THEOREM 2. Equation}
\begin{equation}
\max{ \beta_2 \left(X_1, X_2, ..., X_{n-1} , X_n\right)}  \label{eq:12}
\end{equation}
\textit{could be solved for $\forall X_i \in \{ 0, 1 \}$ in $O\left( n^{3.5} \right)$}.

\textit{Proof}. Let start to investigate \ref{eq:12} when $\forall X_i \in \mathbb{R}, i \in \{ 1, 2, ..., n\}$. According to \ref{eq:10} equation \ref{eq:12} is linear and could be rewritten as follow
\begin{equation}
\begin{split}
&\max{ \mu \left(X_1 + X_2, \mu \left( X_3 + X_4, ... , \mu \left( X_{n-3} + X_{n-2}, X_{n-1} + X_{n} \right) \right) \right)} =  \\
&\mu \left(\max{X_1 + X_2}, \mu \left( \max{X_3 + X_4}, ... , \mu \left( \max{X_{n-3} + X_{n-2}}, \max{X_{n-1} + X_{n}} \right) \right) \right) 
\end{split}
\end{equation}
So we get $m$ equations of $\max{X_{k-1} + X_{k}}$. Equation \ref{eq:12} have solution if system of $m$ equations
\begin{equation}
  \left\{
  \begin{array}{l}
    X_{n-1} + X_{n} \le 2 \ \text{where}\ X_{n-1} \ge 0\ \wedge\ X_{n} \ge 0\\
    \dots  \\
    X_{k-1} + X_{k} \le 2 \ \text{where}\ X_{k-1} \ge 0\ \wedge\ X_{k} \ge 0\\
    \dots  \\
    X_{1} + X_{2} \le 2 \ \text{where}\ X_{1} \ge 0 \wedge\ X_{2} \ge 0
  \end{array} \right. \label{eq:15}
\end{equation}
have solution. This equations for $X_i$ could be solved using best known algorithm of linear programming ~\cite{bib:Karmarkar} in $O\left(n^{3.5}\right)$ and than $\forall \tilde{X}_i \in \{ 0, 1 \}$ expressed as follow 
\begin{equation}
\tilde{X}_i = g^2_0\left( X_i \right), \forall i \in \{ 1, 2, ..., n \} \label{eq:11}
\end{equation}
$\bigcirc$
  
\section{3SAT is in P}
\textit{THEOREM 3. If binary multi-variable logic function is expressed as}
\begin{equation}
\beta_3 \left(X_1, X_2, ... , X_n\right) = (X_1 \vee X_2 \vee X_3) \wedge (X_3 \vee X_4 \vee X_5) \wedge \dots \wedge (X_{n-2} \vee X_{n-1} \vee X_n),
\end{equation}   
\textit{it could be calculated in} $O\left(m\right)$ \textit{where $m$ is number of clauses and $m \ge n$}.

\textit{Proof}. Let start to investigate $\beta$. It could be expressed in notations of LEMMA 4 as
\begin{align}
\mu \left(a,b\right) &= 
  \begin{array}{c|c c c c}
  a \backslash b &  \mbox{$0$} & \mbox{$1$} & \mbox{$2$} & \mbox{$3$}\\  
  \hline  
  \mbox{$0$} & \mbox{$0$} & \mbox{$0$} & \mbox{$0$} & \mbox{$0$}\\
  \mbox{$1$} & \mbox{$0$} & \mbox{$1$} & \mbox{$1$} & \mbox{$1$}\\
  \mbox{$2$} & \mbox{$0$} & \mbox{$1$} & \mbox{$1$} & \mbox{$1$}\\
  \mbox{$3$} & \mbox{$0$} & \mbox{$1$} & \mbox{$1$} & \mbox{$1$}
 \end{array} 
\end{align} 
\begin{equation}
\begin{split} 
 &\beta_3 \left(X_1, X_2, ..., X_{n-1} , X_n\right) = \\
 &\ \mu \left(X_1 + X_2 + X_3, \mu \left( X_4 + X_5 + X_6, ... , \mu \left( X_{n-5} + X_{n-4} + X_{n-3},X_{n-2} + X_{n-1} + X_{n} \right) \right) \right)
\end{split}
\end{equation}
where $+$ is algebraic summation. So expressed $\beta$ function could be calculated  within $2m/3$ summations and $m/3-1$ calls of $\mu$ function. Every result of $\mu$ function (getting item from 2 dimensional array ) could be calculated within $2$ summation operations. So total $\beta$ function calculation time could be expressed as
\begin{align}
T_{\sum} = t_{+} \frac{2 m}{3} + 2 t_{+} \left( \frac{m}{3} -1\right) + p t_{-}= O\left(m\right)
\end{align}  
where $t_+$ is algebraic summation time of two variables and is constant, $t_{-}$ is unary negotiation time and $p$ amount of negotiation functions $p \le n$. $\beta$ function could be tested in linear time. $\bigcirc$

\textit{THEOREM 4. Equation}
\begin{equation}
\max{ \beta_3 \left(X_1, X_2, ..., X_{n-1} , X_n\right)}  \label{eq:20}
\end{equation}
\textit{could be solved for $\forall X_i \in \{ 0, 1\}$ in $O\left( n^{3.5} \right)$}.

\textit{Proof}. Let start to investigate \ref{eq:20} when $\forall X_i \in \mathbb{R}, i \in \{ 1, 2, ..., n\}$. According to \ref{eq:10} equation \ref{eq:20} could be rewritten as follow
\begin{equation}
\begin{split}
&\max{ \mu \left(X_1 + X_2 + X_3, \mu \left( X_4 + X_5 + X_6, ... , \mu \left( X_{n-5} + X_{n-4} + X_{n-3},X_{n-2} + X_{n-1} + X_{n} \right) \right) \right)} =  \\
&\ \mu \left(\max{X_1 + X_2 + X_3}, \mu \left(\max{ X_4 + X_5 + X_6}, ... ,\right. \right. \\
 &\left. \left. \mu \left(\max{ X_{n-5} + X_{n-4} + X_{n-3}},\max{X_{n-2} + X_{n-1} + X_{n}} \right) \right) \right) 
\end{split}
\end{equation}
So we get $m$ equations of $\max{X_{k-2} + X_{k-1} + X_{k}}$. Equation \ref{eq:20} have solution if system of $m$ equations
\begin{equation}
  \left\{
  \begin{array}{l}
    X_{n-2} + X_{n-1} + X_{n} \le 3 \ \text{where}\ X_{n-2} \ge 0\ \wedge X_{n-1} \ge 0\ \wedge\ X_{n} \ge 0\\
    \dots  \\
    X_{k-2} + X_{k-1} + X_{k} \le 3 \ \text{where}\ X_{k-2} \ge 0\ \wedge X_{k-1} \ge 0\ \wedge\ X_{k} \ge 0\\
    \dots  \\
    X_{1} + X_{2} + X_{3}\le 3 \ \text{where}\ X_{1} \ge 0 \wedge\ X_{2} \ge 0 \wedge\ X_{3} \ge 0
  \end{array} \right. \label{eq:21}
\end{equation}
have solution. This equations for $X_i$ could be solved using best known algorithm of linear programming ~\cite{bib:Karmarkar} in $O\left(n^{3.5}\right)$ and than $\forall \tilde{X}_i \in \{ 0, 1 \}$ expressed as follow 
\begin{equation}
\tilde{X}_i = g^2_0\left( X_i \right), \forall i \in \{ 1, 2, ..., n \} \nonumber
\end{equation} $\bigcirc$

\section{kSAT is in P}
\textit{THEOREM 5. If binary multi-variable logic function is expressed as}
\begin{equation}
\beta_k \left(X_1, X_2, ... , X_n\right) = ( \underset{i=1}{\overset{k}{\vee}} X_i ) \wedge (\underset{i=k+1}{\overset{2 k}{\vee}} X_i) \wedge \dots \wedge (\underset{i=m-k+1}{\overset{m}{\vee}} X_i),
\end{equation}   
\textit{it could be calculated in} $O\left(m\right)$ \textit{where $m$ is number of clauses and $m \ge n$}.

\textit{Proof}. Let start to investigate $\beta$. It could be expressed in notations of LEMMA 4 as
\begin{align}
\mu \left(a,b\right) &= 
  \begin{array}{c|c c c c c}
  a \backslash b &  \mbox{$0$} & \mbox{$1$} & \mbox{$2$} & \dots & \mbox{$n-1$}\\  
  \hline  
  \mbox{$0$} & \mbox{$0$} & \mbox{$0$} & \mbox{$0$} & \dots & \mbox{$0$}\\
  \mbox{$1$} & \mbox{$0$} & \mbox{$1$} & \mbox{$1$} & \dots & \mbox{$1$}\\
  \mbox{$2$} & \mbox{$0$} & \mbox{$1$} & \mbox{$1$} & \dots & \mbox{$1$}\\
  \dots & & & & & \\  
  \mbox{$n-1$} & \mbox{$0$} & \mbox{$1$} & \mbox{$1$} & \dots & \mbox{$1$}
 \end{array} 
\end{align} 
\begin{equation}
\begin{split}
 &\beta_k \left(X_1, X_2, ..., X_{n-1} , X_n\right) = \\
 &\mu \left(\sum_{i=1}^{k} X_i, \mu \left( \sum_{i=k+1}^{2 k} X_i, ... , \mu \left( \sum_{i=n-2k+1}^{n-k} X_i,\sum_{i=n-k+1}^{n} X_i \right) \right) \right) \label{eq:18}
\end{split}
\end{equation}
So expressed $\beta$ function could be calculated  within $\left(k-1 \right)m/k$ summations and $m/k-1$ calls of $\mu$ function. Every result of $\mu$ function (getting item from 2 dimensional array ) could be calculated within $2$ summation operations. So total $\beta$ function calculation time could be expressed as
\begin{align}
T_{\sum} = t_{+} \frac{\left(k-1 \right) m}{k} + 2 t_{+} \left( \frac{m}{k} -1\right) + p t_{-}= O\left(m\right)
\end{align}  
where $t_+$ is algebraic summation time of two variables and is constant,  $t_{-}$ is unary negotiation time and $p$ amount of negotiation functions $p \le n$. $\beta$ function could be tested in linear time. $\bigcirc$

\textit{THEOREM 6. Equation}
\begin{equation}
\max{ \beta_k \left(X_1, X_2, ..., X_{n-1} , X_n\right)}  \label{eq:22}
\end{equation}
\textit{could be solved for $\forall X_i \in \{ 0, 1 \}$ in $O\left( n^{3.5} \right)$}.

\textit{Proof}. Let start to investigate \ref{eq:22} when $\forall X_i \in \mathbb{R}, i \in \{ 1, 2, ..., n\}$. According to \ref{eq:10} equation \ref{eq:22} could be rewritten as follow
\begin{equation}
\begin{split}
&\max{ \mu \left(\sum_{i=1}^{k} X_i, \mu \left( \sum_{i=k+1}^{2 k} X_i, ... , \mu \left( \sum_{i=n-2k+1}^{n-k} X_i,\sum_{i=n-k+1}^{n} X_i \right) \right) \right)} =  \\
&\mu \left(\max \sum_{i=1}^{k} X_i, \mu \left( \max \sum_{i=k+1}^{2 k} X_i, ... , \mu \left(\max \sum_{i=n-2k+1}^{n-k} X_i,\max \sum_{i=n-k+1}^{n} X_i \right) \right) \right) 
\end{split}
\end{equation}
So we get $m$ equations of $\max \sum_{i=k+1}^{2 k} X_i$. Equation \ref{eq:22} have solution if system of $m$ equations
\begin{equation}
  \left\{
  \begin{array}{l}
    \sum_{i=n-k+1}^{n} X_i \le k \ \\
    \dots  \\
    \sum_{i=k+1}^{2 k} X_i \le k \ \text{where}\ X_{i} \ge 0\ \wedge \forall i \in \{1, 2, ..., n\} \\
    \dots  \\
    \sum_{i=1}^{k} X_i\le k \ 
  \end{array} \right. \label{eq:23}
\end{equation}
have solution. This equations for $Y_i$ could be solved using best known algorithm of linear programming ~\cite{bib:Karmarkar} in $O\left(n^{3.5}\right)$ and than $\forall \tilde{X}_i \in \{ 0, 1 \}$ expressed as follow 
\begin{equation}
\tilde{X}_i = g^k_0\left( X_i \right), \forall i \in \{ 1, 2, ..., n \} \nonumber
\end{equation} $\bigcirc$

\end{document}